\documentclass{ws-rv9x6}
\usepackage[square]{ws-rv-van}             
\makeindex

\def\grad{{\mbox{\boldmath$\nabla$\unboldmath}}}

\begin{document}

\chapter[Two-Fluid Model of Superfluidity]{From Classical Fields to
Two-Fluid Model of Superfluidity:\\
Emergent Kinetics and Local Gauge Transformations \label{ch1}}

\author[H.\ Salman]{Hayder Salman}

\address{School of Mathematics, University of East Anglia,
Norwich, NR4 7TJ, UK} 

\author[N.G.\ Berloff]{Natalia G.\ Berloff} 

\address{DAMTP, University of Cambridge, Cambridge, CB3 0WA, UK}

\author[H.\ Salman, N.G.\ Berloff, and P.H.\ Roberts]{Paul H.\ Roberts} 

\address{Department of Mathematics, University of California Los Angeles,\\
Los Angeles, CA 90095-1555, USA}

\begin{abstract}
The first successful macroscopic theory for the motion of superfluid
helium was that of Lev Landau (1941) in which the fluid is
modelled phenomenologically as an interpenetrating mixture of a superfluid and a normal fluid. It has
later been shown that Landau's two-fluid model can be rigorously derived
from a one-fluid model within the classical fields approximation.
Assuming a separation of scales exists between
the slowly varying, large-scale, background (condensate) field, and the short
rapidly evolving excitations, a full description of the kinetics between the condensate and the thermal cloud can be obtained. The kinetics describes three-wave and four-wave interactions that resemble
the $C_{12}$ and $C_{22}$ terms, respectively, in the collision
integral of the ZNG theory. The scale separation assumption
precludes analysis of the healing layer and thus does not include
the dynamics of quantised vortices. Whilst the analysis
required the use of small parameters arising from the scale separation
assumption and the assumption of a weakly depleted condensate, we expect the results to hold true
over a wider range of parameters. This belief is motivated by the
validity of Landau's two-fluid model which
can be derived from a one-fluid model using nothing more than Galilean
invariance principles. Indeed, we argue that similar arguments can be
used to recover a two-fluid model directly from a classical field
simply by invoking a local gauge transformation. This derivation does
not require any small parameters to be introduced suggesting that the
results that lead to the kinetic equations may turn out to be more
general. 
\end{abstract}

\body

\section{Introduction}

Historically, the development of the two-fluid model of superfluidity for liquid He II has relied on a phenomenological approach. By starting with the governing equations of motion for a single fluid, and introducing an additional velocity field associated with the superfluid component, Landau \cite{landau_41,landau_lifshitz_87} formulated his seminal
two-fluid model of liquid He II using little more than the 
principle of Galilean invariance. Landau's model predated the
discovery of quantized vortices which were conjectured more than a decade 
later by Onsager \cite{onsager_49} and Feynman \cite{gorter_book_55}. These conjectures received further support with the theories of Gross \cite{gross_61} and Pitaevskii \cite{pitaevskii_61} formulated for a weakly interacting Bose gas that were derived from microscopic considerations.

Some authors, notably Hills and Roberts \cite{hills_roberts_77}, Ginzburg and Pitaevskii \cite{ginzburg_pitaevskii_58}, Khalatnikov \cite{khalatnikov_70}, and Geurst \cite{geurst_80},  have sought
generalizations of Landau's two-fluid model by including relaxation effects
as well as the effects of healing. The latter is particularly important in light of the presence of quantized vortices. In fact, it is well established that the Euler equations, governing the motion of an ideal classical fluid, preserve the topology imprinted by the initial conditions onto the flow. In contrast, vortices in superfluids undergo reconnections (e.g.\ \cite{koplik_levine_93}) which consequently have important effects on their dynamics \cite{schwarz_85}. These can only occur if the effects of healing are retained in the model.  
The rationale for including the effects of healing into Landau's two-fluid model is, therefore, perfectly understandable.
However, the validity of such a two-fluid description for Helium II per se has been brought into question (Ginzburg and Sobyanin \cite{ginzburg_sobyanin_76}). This is due to the healing length in He II being of the order of the inter-atomic distances thereby precluding a continuum description on these scales.

To account for the effects of healing and relaxation in a physically consistent manner, we will review a more fundamental derivation of the two-fluid model
first proposed by Putterman and Roberts \cite{putterman_roberts_82, putterman_roberts_83} (hereafter PR). PR arrived at the two-fluid model through an intermediate step involving the derivation of a kinetic set of equations. Their work is in some sense the basis of the classical fields approach. This was developed further by Kagan and Svistunov \cite{kagan_svistunov_97}, and Davis {\em et al.} \cite{davis_morgan_burnett_02}, as a model for studying finite-temperature effects in atomic Bose-Einstein condensates. 
Our primary goal in this chapter is, therefore, to review some of
these results and to establish their connection to other closely related theories. In particular, we will show that the local-gauge transformation used by Coste \cite{coste_98} is a limiting case of the PR theory.

\section{Kinetics in Nonlinear Classical Fields \label{sec_kinetics}} 

In recent years, the classical fields approximation has evolved from a
qualitative method to describe the kinetics of condensate formation from a
strongly nonequilibrium initial condition (Berloff and Svistunov \cite{berloff_svistunov_02}), to a method that is used to model finite temperature Bose gases \cite{brewczyk_gajda_rzazewski_07}. In the context of modelling superfluid He II, PR were two of the earliest advocates of the classical fields approximation. The key objective of these authors at the time was to elucidate the kinetics that ultimately give rise to a two-fluid
description of liquid Helium II. Remarkably, by starting from the equation of
a single fluid, and using a scale-separation argument, they were able to recover
a set of kinetic equations governing the thermal excitations. In the collision-dominated regime this leads to the two-fluid model of Landau.

Following Putterman and Roberts \cite{putterman_roberts_91}, we will use the GP equation in the classical fields approximation as our starting point. This will facilitate in relating the results presented here to the application of local gauge transformations. The GP equation for the classical field $\psi(\mathbf{x},t)$ is
\begin{eqnarray}
i\hbar \psi_t = -\frac{\hbar^2}{2m}\grad^2 \psi + g |\psi^2| \psi.
\end{eqnarray}
If we define $\psi(\mathbf{x},t)=|\psi(\mathbf{x},t)|
\exp[i\varphi(\mathbf{x},t)m/\hbar]$, where $\rho(\mathbf{x},t)=m|\psi|^2$, and $\mathbf{u}(\mathbf{x},t)=\grad \varphi$, this equation can be written in hydrodynamic form as
\begin{eqnarray}
\rho_t + \grad \cdot (\rho \mathbf{u}) = 0, \;\; \varphi_t + \frac{1}{2} \mathbf{u}^2 + \left( \frac{g\rho}{m^2} - \frac{\hbar^2}{2m^2} \frac{\grad^2 \rho^{1/2}}{\rho^{1/2}} \right) =0. \;\;\;\;
\label{salman_eqn_GP_hydro}
\end{eqnarray}
At this point, we wish to stress that the hydrodynamic form given by Eqs.\ (\ref{salman_eqn_GP_hydro}) must not be attributed to the superfluid
density and the superfluid velocity. All we have done is simply to
re-express the equation for the classical field in terms of an amplitude
and a phase.

To proceed, we will consider small amplitude excitations
on top of a background field. We do this by {\em assuming} that the system contains two disparate length and time scales.
We can then decompose the field variables into
a long-wavelength  slowly varying background field denoted with subscript `0', and short wavelength rapidly varying excitations, denoted by primed quantities.
Formally, we introduce a slow timescale $\tau$ and a long-wavelength scale
$\mathbf{X}$ such that $\tau = t \delta$ and $\mathbf{X} = \mathbf{x} \delta$.
Then we have 
\begin{eqnarray}
\!\!\!\!\!\!\!\!\!\!\!\!\! \varphi(\mathbf{x},t) = \varphi_0(\mathbf{X},\tau) + \varphi'(\mathbf{X},\tau,\mathbf{x},t),
\;\;  
\rho(\mathbf{x},t)    = \rho_0(\mathbf{X},\tau) + \rho'(\mathbf{X},\tau,\mathbf{x},t).
\label{salman_eqn_perturb} 
\end{eqnarray}
The small amplitude requirement is enforced by requiring $|\varphi'/\varphi_0| \sim \mathcal{O}(\epsilon) \ll 1$ and likewise for $|\rho'/\rho_0|$. Moreover, given the scale separation assumption, we will
decompose the perturbation into 
\begin{eqnarray}
\!\!\!\!\!\!\!\!\!\! \varphi' = \epsilon \left[ \varphi_1(\mathbf{X},\tau) + 
\delta \varphi_2(\mathbf{X},\tau) \right] \mathrm{e}^{i\Theta}, \;\;\;
\rho' = \epsilon \left[ \rho_1(\mathbf{X},\tau) + 
\delta \rho_2(\mathbf{X},\tau) \right] \mathrm{e}^{i\Theta}.
\label{salman_eqn_scaling}
\end{eqnarray}
The perturbation is, therefore, represented by a monochromatic wave with a varying wavelength controlled by the parameter $\delta \ll 1$. With $\mathrm{d}\Theta(\mathbf{x},t) = \mathbf{k} \cdot \mathrm{d}\mathbf{x} - \omega \mathrm{d}t$ 
corresponding to the rapidly varying scales, we have,
\begin{eqnarray}
\omega(\mathbf{X},\tau) = -\partial \Theta/\partial t, \;\;\;\; \mathbf{k}(\mathbf{X},\tau) = \grad \Theta.
\end{eqnarray}
We shall begin by assuming that $\epsilon^2 \ll \delta \ll \epsilon \ll 1$. 
We then seek how excitations propagate on some initial background state denoted by $\rho(\mathbf{X},0)=\rho_0(\mathbf{X})$ and 
$\mathbf{v}(\mathbf{X},0)=\mathbf{v}_0(\mathbf{X})$. Substituting Eqs.\
(\ref{salman_eqn_perturb}), (\ref{salman_eqn_scaling}) into Eq.\ (\ref{salman_eqn_GP_hydro}) and expanding to $\mathcal{O}(\epsilon)$, we obtain the local dispersion relationship
\begin{eqnarray}
\omega = \overline{\omega} + \mathbf{k} \cdot \mathbf{v}_0, \;\;\;\;\;\;
\overline{\omega}_{\pm} = \pm c_0 \mathbf{|k|} = \pm c_0 k, \;\;\;\;\;\; 
c_0 = (\sqrt{g \rho_0})/m.
\end{eqnarray}
where $\overline{\omega}$ is the intrinsic frequency, (the frequency in the frame of reference of $\mathbf{v}_0$) and $c_0$ is the speed of sound.
At $\mathcal{O}(\epsilon \delta)$, we obtain a conservation law 
\begin{eqnarray}
\frac{\partial \mathcal{A}}{\partial t} + \frac{\partial}{\partial X_i} (c_{gi}
\mathcal{A}) = 0, \label{salman_eqn_action}
\end{eqnarray}
for the waveaction per unit volume defined by 
$\mathcal{A} = \mathcal{E}/\overline{\omega}$, 
$\mathcal{E} = c_0^2 |\rho_1|^2/\rho_0 = \rho_0 k^2 |\varphi_1|^2$,
where $c_{gi} = \partial \omega/\partial k_i$ is the group velocity.
This conservation law has been obtained by considering a monochromatic, slowly varying wavetrain, of characteristic wavenumber $\mathbf{k} \equiv \mathbf{k}(\mathbf{X},\tau)$. A general initial perturbation, $\rho_1(\mathbf{x},0)$ say, can be represented by a linear superposition of such wavetrains which evolves into $\rho_1(\mathbf{x},t)$ at time $t$. 
In view of the scale separation that is assumed to exist in our system, we introduce the Gabor transformed quantities, such that
\begin{eqnarray}
B(\mathbf{X},\mathbf{k}_0,0) = \int_{-\infty}^{\infty} \!\!\!\!\! \rho'(\mathbf{x}',t)
\mathrm{e}^{i\mathbf{k}_0\cdot(\mathbf{x}-\mathbf{x}')} \mathcal{F}(|\mathbf{x}-\mathbf{x}'|)
d\mathbf{x}',
\end{eqnarray}  
with $V(\mathbf{X},\mathbf{k},t)$ defined in an analogous way from $\varphi'$. 
Following \cite{lvov_nazarenko_03}, we take $\mathcal{F}(\xi)= (\delta^*/\sqrt{2\pi})^3\exp\{-(\delta^*\xi)^2/2\}$ and set $\delta \ll \delta^* \ll 1$ so that the kernel $\mathcal{F}$ varies on the intermediate scale. The inverse transform is
\begin{eqnarray}
\!\!\!\!\!\!\!\!\!\!\!\!\!\!\!\! \rho_1(\mathbf{x},0) &=& \frac{1}{(\sqrt{2\pi}\delta^*)^3} \int_{-\infty}^{\infty} \!\!\!\!\! B(\mathbf{k}_0;\mathbf{X},0)\mathrm{e}^{i{\mathbf{k}_0}\cdot
\mathbf{x}}d\mathbf{k}_0, \;\; 
\mathbf{k}({\mathbf{k}_0;\mathbf{X},0}) = \mathbf{k}_0, \\
\!\!\!\!\!\!\!\!\!\!\!\!\!\!\!\! \rho_1(\mathbf{x},t) &=& \frac{1}{(\sqrt{2\pi}\delta^*)^3} \int_{-\infty}^{\infty} \!\!\!\!\! B(\mathbf{k};\mathbf{X},\tau)
\mathrm{e}^{i\Theta(\mathbf{k};\mathbf{x},t)}d\mathbf{k}_0, \;\;
\mathbf{k}({\mathbf{k}_0;\mathbf{X},\tau}) = \grad \Theta.
\end{eqnarray}
The mean waveaction per unit volume can now be defined as
\begin{eqnarray}
\!\!\!\!\!\!\!\!\!\!\!\! 
\mathcal{A}(\mathbf{k};\mathbf{X},\tau) = (c_0/\rho_0 k) 
\left< |B(\mathbf{k;\mathbf{X},\tau})|^2 \right>,
\end{eqnarray}
with the waveaction density per unit volume given by the integral of $\mathcal{A}(\mathbf{k};\mathbf{X},\tau)$ with respect to the initial wavenumbers $\mathbf{k}_0$. To 
express this integral over the wavenumbers $\mathbf{k}$, we introduce $n=\sigma \mathcal{A}$ where $\sigma$ denotes the Jacobian $\partial \mathbf{k}_0/\partial \mathbf{k}$. This Jacobian satisfies Liouville's theorem (Soward \cite{soward_75})
\begin{eqnarray}
\frac{\partial \sigma}{\partial \tau} + c_{gi} \frac{\partial \sigma}{\partial X_i} = \sigma \frac{\partial c_{gi}}{\partial X_i}. 
\end{eqnarray}
Combining the above Eqn.\ with Eqn.\ (\ref{salman_eqn_action}) and then using the chain rule to transform from the independent variables $(\mathbf{k}_0,\mathbf{X},\tau)$ that is used in Eq.\ (\ref{salman_eqn_action}) to the variables $(\mathbf{k},\mathbf{X},\tau)$, we finally arrive at the wave Vlasov equation
\begin{eqnarray}
\frac{\partial n}{\partial \tau} + c_{gi} \frac{\partial n}{\partial X_i} 
- \frac{\partial \omega}{\partial X_i} 
\frac{\partial n}{\partial k_i} = 0.
\end{eqnarray}
As is clear from the form of the equation, this governs the motion of the excitations in the collisionless regime. We note that 
$n(\mathbf{k},\mathbf{x},t)$ can be identified with Bogoliubov quasiparticles (see e.g.\ Lvov {\em et al.} \cite{lvov_nazarenko_03}).

Now in the opposite regime, where we have $\delta \ll \delta^* \ll \epsilon^4 \ll 1$, we obtain, at leading order, a kinetic equation with a collision integral arising from the nonlinear terms. Under these conditions, the kinetic equation can be derived in a {\em rigorous} way using the methods of matched asymptotics. Since these derivations are quite involved, we will merely sketch out the initial key steps and defer the details of the closures used to the relevant references. 
In what follows, we will suppress the dependence on $\mathbf{X}$ since it does not play a key role in the kinetics in the parameter regime we are considering here. It then follows from Eq.\ (\ref{salman_eqn_GP_hydro}) and the above definitions that to $\mathcal{O}(\epsilon^2)$ we have
\begin{eqnarray}
\left( \frac{\partial}{\partial t} + i\mathbf{k} \cdot \mathbf{v}_0 \right)B(\mathbf{k})
&-& \rho k^2 V(\mathbf{k}) = \epsilon \int_{_{\mathbf{k}_1+\mathbf{k}_2  =\mathbf{k}}} \!\!\!\!\!\!\!\!\!\!\!\!\!\!\!\! \mathbf{k} \cdot \mathbf{k}_2
B(\mathbf{k}_1) V(\mathbf{k}_2) d\mathbf{k}_1. \;\;\;\;\;\;\; \label{salman_eqn_Ak} 
\end{eqnarray}
Similarly, we can derive an equation governing $V(\mathbf{k})$. By introducing
\begin{eqnarray}
B(\mathbf{k},t) = \sum_s a^s(\mathbf{k},t)e^{-i\omega_st}, \;\;\;\;\;\; 
\sum_s (\partial a^s/\partial t) e^{-i\omega_st} = 0,
\end{eqnarray}
we can derive from Eq.\ (\ref{salman_eqn_Ak}) and the corresponding equation for $V(\mathbf{k})$ an equation for the evolution of the Fourier amplitudes $a^s(\mathbf{k},t)$ given by
\begin{eqnarray}
\!\!\!\!\!\!\!\!\!\!\!\!\! \frac{\partial a^s(\mathbf{k})}{\partial t} = \frac{\epsilon k}{2isc} \sum_{s_1,s_2}
\int_{_{\mathbf{k}_1+\mathbf{k}_2  =\mathbf{k}}} 
\!\!\!\!\!\!\!\!\!\!\!\!\!\!\!\! H(\mathbf{k}_1,\mathbf{k}_2) 
a^{s_1}(\mathbf{k}_1)a^{s_2}(\mathbf{k_2}) \mathrm{e}^{i(s\overline{\omega}
-s_1\overline{\omega}_1-s_2\overline{\omega}_2)t} d\mathbf{k}_1.
\end{eqnarray}
The coefficient $H$ is defined in PR. Starting from the above equation, we can derive equations for the second order correlation which would correspond to the waveaction density $n(\mathbf{k})$. However, in so doing, we find that it depends on the fourth order correlation that arises in the integrand. The equation for this fourth order correlation in turn depends on even higher order correlations and so the equations cannot be closed. Benney and Saffman \cite{benney_saffman_66} (see also derivations by Connaughton and Pomeau \cite{connaughton_pomeau_04}) show how the hierarchy can be closed under an appropriate set of assumptions (e.g.\ quasi-Gaussian statistics). Further details of the derivation can be found in PR but it can be shown that we finally obtain the kinetic equation
\begin{eqnarray}
\left( \frac{\partial n(\mathbf{k}_0)}{\partial t} \right)_{\mathrm{coll}} = && \epsilon^2 \pi \rho
\int \!\!\!\! \int |H(\mathbf{k}_1,\mathbf{k}_2)|^2 \left\{ n(\mathbf{k}_1)n(\mathbf{k}_2) \right. \label{salman_eqn_C12} \\
&&\left. -n(\mathbf{k}_0)[n(\mathbf{k}_1)+n(\mathbf{k}_2)] \right\}
\delta(\overline{\omega}_0-\overline{\omega}_1-\overline{\omega}_2)
\delta(\mathbf{k}_0-\mathbf{k}_1-\mathbf{k}_2) \nonumber \\
&&-2|H(-\mathbf{k}_1,\mathbf{k}_2)|^2 \left\{ n(\mathbf{k}_0)n(\mathbf{k}_1)
-n(\mathbf{k}_2)[n(\mathbf{k}_0)+n(\mathbf{k}_1)] \right\} \nonumber \\
&&\times \delta(\overline{\omega}_2-\overline{\omega}_0-\overline{\omega}_1)
\delta(\mathbf{k}_2-\mathbf{k}_0-\mathbf{k}_1)
\left( \frac{k_0 k_1 k_2}{c_0 c_1 c_2} \right) d\mathbf{k}_1 d\mathbf{k}_2. \nonumber
\end{eqnarray}
If we extend the above analysis to the next order in $\epsilon$, then we find
\begin{eqnarray}
\left( \frac{\partial n(\mathbf{k}_0)}{\partial t} \right)_{\mathrm{coll}} = && \frac{9\pi \rho^2 \epsilon^4}{2}
\int \!\!\!\! \int \!\!\!\! \int |K(\mathbf{k}_1,-\mathbf{k}_2,-\mathbf{k}_3)|^2 \left\{ n(\mathbf{k}_2)n(\mathbf{k}_3) \right. \label{salman_eqn_C22} \\
&&\times \left. [n(\mathbf{k}_0)+n(\mathbf{k}_1)]
- n(\mathbf{k}_0)n(\mathbf{k}_1)[n(\mathbf{k}_2)+n(\mathbf{k}_3)]  \right\} \nonumber \\
&&\times \delta(\overline{\omega}_2+\overline{\omega}_3-\overline{\omega}_0
-\overline{\omega}_1)
\delta(\mathbf{k}_2+\mathbf{k}_3-\mathbf{k}_0-\mathbf{k}_1) d\mathbf{k}_1  d\mathbf{k}_2. \nonumber
\end{eqnarray}
The coefficient $K$ is defined in PR. The first term proportional to $\epsilon^2$ in Eq.\ (\ref{salman_eqn_C12}) represents resonant triad interactions of the waves with wavenumbers $\mathbf{k}_0, \mathbf{k}_1, \mathbf{k}_2$. The resonance conditions contained in the delta functions arise from conservation of momentum and conservation of energy of the governing equations. We note that this leading order behaviour of the wave kinetics, derived for a condensate containing a large fraction of the total number of particles, has a similar structure to the $C_{12}$ term in the collision integral of the Zaremba, Nikuni, and Griffin (ZNG) model \cite{zaremba_nikuni_99}. On the other hand, the four-wave resonant interactions, described by Eq.\ (\ref{salman_eqn_C22}), corresponds to the $C_{22}$ term in the ZNG theory and represents interactions of quasiparticles within the the thermal cloud. In fact, the key difference between the classical field and ZNG model stems from the missing spontaneous scattering contributions that are retained in the ZNG theory. Given that we have started from a mean classical field representation which assumes macroscopically occupied modes ($n(\mathbf{k}) \gg 1$), it is natural that a classical field cannot model spontaneous scattering processes.

Using the above results, we can now recover a hydrodynamic two-fluid model. To help identify the microscopic basis of the results to be derived in the next section, using local gauge transformations, we have opted to present a brief derivation. We recall that in the collisionless regime, corresponding to $\epsilon^2 \ll \delta \ll \epsilon \ll 1$, we obtained the Vlasov wave equation. On the other hand, the kinetic equation was obtained in the collision-dominated regime when $\delta \ll \epsilon^4 \ll 1$. If we relax our condition on $\delta$ by requiring $\epsilon, \delta \ll 1$, then we expect to obtain a Boltzmann like equation of the form
\begin{eqnarray}
\frac{\partial n}{\partial t} + \mathbf{c}_g \cdot \grad n 
- \grad \omega \cdot \grad_k n = \left( \frac{\partial n}{\partial t} \right)_{\mathrm{coll}}.
\end{eqnarray} 
From this equation, we can show that 
\begin{eqnarray}
&&\frac{\partial}{\partial t} \int n \omega d\mathbf{k} + \nabla_j \int n \omega c_{gj} d\mathbf{k} = \int n \frac{\partial \omega}{\partial t} d\mathbf{k}, \\
&&\frac{\partial}{\partial t} \int n k_i d\mathbf{k} + \nabla_j \int n k_i 
c_{gj} d\mathbf{k} = -\int n \frac{\partial \omega}{\partial x_i} d\mathbf{k}, \label{salman_eqn_mom_dens} \\
&&\frac{\partial}{\partial t} \int \Sigma(n) d\mathbf{k} + \nabla_j \int \Sigma(n) c_{gj} d\mathbf{k} \ge 0, \label{salman_eqn_entropy_dens}
\end{eqnarray}
where $\Sigma(n)$ is the entropy density. In the collision-dominated regime, we have local thermodynamic equilibrium with $(\partial n/\partial t)_{\mathrm{coll}}=0$. The equilibrium distribution 
can then be written in terms of the temperature $T$ as  
\begin{eqnarray}
n = n_{\mathrm{eq}}(\beta \Omega), \;\;\;\;\; \beta=1/k_B T, \;\;\;\;\;
\Omega = \overline{\omega} - \mathbf{k} \cdot \mathbf{w}.
\end{eqnarray}
We note that the equilibrium Rayleigh-Jeans distribution given above is the long-wavelength (classical field) limit of the Bose-Einstein distribution. We also note that $n=n_{\mathrm{eq}}$ is the only solution for which equality is obtained in Eq.\ (\ref{salman_eqn_entropy_dens}). This follows from the H-theorem corresponding to the case where the entropy is maximised. Now using $n_{\mathrm{eq}}$, we can define the following thermodynamic variables at each point $\mathbf{X}$,
\begin{eqnarray}
E_n = \int n_{\mathrm{eq}} \overline{\omega} d\mathbf{k}, \;\;\;\; 
\rho_n \mathbf{w} = \int n_{\mathrm{eq}} \mathbf{k} d\mathbf{k}, \;\;\;\;
S = \int \Sigma(n_{\mathrm{eq}}) d\mathbf{k}.
\end{eqnarray}
Hence, from Eq.\ (\ref{salman_eqn_mom_dens}) and (\ref{salman_eqn_entropy_dens}), we obtain
\begin{eqnarray}
\frac{\partial}{\partial t}(\rho_n w_i) + \frac{\partial}{\partial x_j}
(\rho_n w_i v_{nj}) &=& -S \frac{\partial T}{\partial x_i} - \rho_n w_j 
\frac{\partial v_{nj}}{\partial x_i} \\
\frac{\partial S}{\partial t} + \grad \cdot (S\mathbf{v}_n) &=& 0,
\end{eqnarray}
where the entropy per unit volume ($S$) is advected by the normal fluid component. To obtain the equation of continuity and superfluid velocity, we expand the original equations of motion up to $\mathcal{O}(\epsilon^2)$ so that
\begin{eqnarray}
\frac{\partial \rho_0}{\partial t} + \grad \cdot (\rho_0 \mathbf{v}_0 
+ \rho_n \mathbf{w}) = 0, \;\;\;\;\;\;\;\;
\frac{\partial \mathbf{v}_0}{\partial t} + \mathbf{v}_0 \cdot \grad \mathbf{v}_0 = - \grad \mu,
\end{eqnarray}
where $\mu$ is the chemical potential and is related to the pressure $P$ by \cite{putterman_roberts_91}
\begin{eqnarray}
\mu = \Phi_0 + \frac{E_n}{c_0} \left( \frac{dc_0}{d\rho_0} \right), \;\;\;\; \Phi_0 = \int^{\rho_0} \frac{dP}{\rho}.
\end{eqnarray}
Attributing $\rho_0, \mathbf{v}_0, \mathbf{w}$ to the total density $\rho$, the superfluid velocity $\mathbf{v}_s$ and the relative normal/superfluid velocity $(\mathbf{v}_n-\mathbf{v}_s)$, respectively, we recover Landau's two-fluid model. We note that one can include dissipative effects by
extending the above analysis using a Chapman-Enskog \cite{chapman_cowling_52} procedure. 

We end this section by asking why two-fluid phenomena, whose premise is a one-fluid model, cannot be observed in classical fluids. It turns out that the reason for this is set by the requirement $\rho_n < \rho$. This condition is typically satisfied only by He II at very low temperatures thus preventing this phenomena from being observed in other classical fluids. 

\section{Two-Fluid Model from Local Gauge Transformations}

Akin to Landau's original derivation of the two-fluid model from a one-fluid model, PR have revealed the kinetic basis of the two-fluid theory and how it emerges from a one-fluid model. In that sense, their results generalise Landau's result since a kinetic description remains valid even when a hydrodynamic description breaks down. This occurs, for example, very close to $T=0$ where the mean free path of the excitations becomes so large that a hydrodynamic description of the normal fluid component no longer holds. However, one can also arrive at the two-fluid model directly from the GP equation. To show this, we recall that the GP equation can be derived from an action principle with the Lagrangian density given by
\begin{eqnarray}
\mathcal{L}_0(\psi,\psi^*) = \frac{i\hbar}{2} \left[ \psi(\partial_t \psi)^* 
-\psi^* (\partial_t \psi) \right] + \frac{\hbar^2}{2m} |\grad \psi|^2 + \frac{g}{2}|\psi|^4. \label{salman_eqn_Lag0}
\end{eqnarray}
The above Lagrangian density is invariant under the global gauge transformation
$\psi \rightarrow \psi \mathrm{e}^{i(\alpha m/\hbar)}$ ($\alpha$ is a real constant) and under spatial and time translations of the form $\mathbf{x} \rightarrow \mathbf{x} + \delta\mathbf{x}$ and $t \rightarrow t + \delta t$. It follows from Noether's theorem that these symmetries lead to the conservation laws of mass conservation, momentum conservation, and energy conservation, respectively. Since two-fluid hydrodynamics is a manifestation of an additional macroscopic degree of freedom, we expect the respective conservation law can be attributed to another symmetry in the Lagrangian density. We argue that the new macroscopic degree of freedom is associated with a broken local gauge symmetry. 
Early attempts to derive two-fluid hydrodynamics from local gauge transformations were carried out by Chela-Flores \cite{chela_flores_75}, and Cummings {\em et al.} \cite{cummings_herold_70}. However, the resulting equations were inconsistent with Landau's model since they did not respect Galilean invariance. Coste \cite{coste_98} presented an alternative derivation of Landau's two-fluid model using local gauge transformations. Coste proceeded in a phenomenological way resting his ideas on the assumption that the GP equation only describes the condensate close to $T=0$. However, we argue that the approach does in fact have a more fundamental basis arising as a limiting case from the kinetic description presented in the previous section. In fact, in the limit where local thermodynamic equilibrium holds, the macroscopic variables, such as superfluid and normal fluid velocities, become functions of space and time. The local gauge transformation essentially involves introducing a local gauge field that is related to these variables as we will demonstrate.

We begin by noting that the GP equation is invariant under a Galilean transformation of the form $\mathbf{x} \rightarrow \mathbf{x} + \mathbf{V} t$, and $t \rightarrow t$ provided the wavefunction is transformed according to 
$\psi(\mathbf{x},t) \rightarrow \psi(\mathbf{x},t)\exp[i(-\mathbf{V} \cdot \mathbf{x} + \frac{1}{2}V^2t)m/\hbar]$. Rewriting the field $\psi(\mathbf{x},t)$ in terms of an amplitude $|\psi|$ and a phase ($\varphi m/\hbar$), we have
$\grad \varphi \rightarrow \grad \varphi + \mathbf{V}$. The transformation, therefore, relates $\psi$ in one frame of reference to $\psi$ in a frame of reference moving with velocity $\mathbf{V}$. If we demand that $\grad \varphi$ correspond to the local superfluid velocity, then we must transform to a local frame of reference such that this requirement is satisfied. This is readily achieved by introducing a local gauge transformation 
$\psi \rightarrow \psi \mathrm{e}^{i\alpha(\mathbf{x},t)m/\hbar}$.
If we now define $\zeta \equiv -\partial_t \alpha$ and $\mathbf{A} \equiv \grad \alpha$, we can determine the forms of $\zeta$ and $\mathbf{A}$ by noting that for the Galilean transformation given above, we have $\grad \rightarrow \grad$ and 
$\partial_t \rightarrow \partial_t - \mathbf{V} \cdot \grad$. Since 
the spatial derivative does not change under a Galilean boost, and $\mathbf{A}$ is defined as the spatial derivative of $\alpha$, it follows that 
$\mathbf{A}$ must also remain invariant.
Similarly, using the transformation of the time derivative with $\mathbf{V}$  coinciding with the local normal fluid velocity $\mathbf{v}_n$, it follows that
\begin{eqnarray}
\mathbf{A} = \chi(\rho,S) (\mathbf{v}_s-\mathbf{v}_n), \;\;\;\; 
\zeta = \eta(\rho,S) + \mathbf{v}_n \cdot \mathbf{A}.
\end{eqnarray}
$\chi$ and $\eta$ are Galilean invariant scalars that are functions of the density and entropy only. Now under the local gauge transformation, the Lagrangian density given in Eq.\ (\ref{salman_eqn_Lag0}) transforms as
\begin{eqnarray}
\mathcal{L}_1(\psi,\psi^*,S,\mathbf{v}_n) &=& \mathcal{L}_0 + \frac{\hbar}{2i} \left( \psi^* \grad \psi  
-\psi \grad \psi^* \right) \cdot \mathbf{A} + \frac{m}{2} A^2 |\psi|^2 \nonumber - m|\psi|^2 \zeta \\
&=& \rho \partial_t \varphi + \frac{g \rho^2}{2m^2} + \frac{\rho}{2} 
(\grad \varphi)^2 + \frac{\hbar^2}{2m^2} \frac{(\grad \rho)^2}{4\rho} - \rho \zeta \nonumber \\
&+& \frac{\rho}{2}(\chi^2 - 2\chi)v_s^2 + \rho \chi (1-\chi) \mathbf{v}_n \cdot
\mathbf{v}_s + \frac{\rho}{2} \chi^2 v_n^2 . \label{salman_eqn_Lagrangian1}
\end{eqnarray}
The Euler-Lagrange equation for $\varphi$ now reads
\begin{eqnarray}
\frac{\partial \rho}{\partial t} + \grad \cdot \left[ \rho(1-\chi)^2 
\mathbf{v}_s + \rho \chi (1-\chi) \mathbf{v}_n \right] = 0. 
\label{salman_eqn_mass_conv} 
\end{eqnarray}
Equation (\ref{salman_eqn_mass_conv}) becomes the two-fluid equation for mass conservation when
\begin{eqnarray}
\rho_n = \rho \chi(\rho,S) [2-\chi (\rho,S)], \;\;\;
\rho_s = \rho [1-\chi (\rho,S)]^2.
\end{eqnarray}
Similarly, the Euler-Lagrange equation for $\rho$ gives
\begin{eqnarray}
\!\!\!\!\!\!\!\!\!\!\!\!\!
\frac{\partial \varphi}{\partial t} &+& \frac{1}{2} (\grad \varphi)^2 + \mu = 
\frac{\hbar^2}{2m^2}\left[ \frac{\grad^2 \rho}{4\rho} - \frac{(\grad \rho)^2}{8\rho^2} \right], \nonumber \\
\!\!\!\!\!\!\!\!\!\!\!\!\!
\mu \equiv \eta + \rho \frac{\partial \eta}{\partial \rho} &+& \frac{g\rho}{m^2}
- \frac{1}{2} \left[ \frac{1}{2} \chi (2-\chi) + \rho (1-\chi) \frac{\partial \chi}{\partial \rho} \right] |\mathbf{v}_s-\mathbf{v}_n|^2.
\end{eqnarray}
These expressions show that $\chi \rightarrow 0$, and
$\rho \rightarrow \rho_s$, as $T \rightarrow 0$.
It then follows that the generalised Lagrangian density reduces to the original GP equation that governs the condensate motion close to $T=0$ when $\zeta \rightarrow \mu$.
Thus far, we have recovered the equation for mass conservation,and the equation for the superfluid velocity. To obtain the remaining set of equations for the two-fluid model, we need to introduce additional constraints on the system through the use of Clebsch potentials in line with Geurst's \cite{geurst_80} variational formulation. These additional constraints are required since it has been known for some time \cite{Eckart_38,Herivel_55} that the equations of motion of hydrodynamics, when derived from a variational principle, tend to be constrained to an irrotational flow. This is associated with a particle relabelling symmetry that is lost in the Eulerian representation of a fluid and must be enforced as a separate constraint through the use of Lagrange multipliers. The distinguishability of a `fluid element' is associated with the thermal excitations that satisfy an independent equilibrium distribution at each point in space. The final form of the Lagrangian density is therefore given by 
\begin{eqnarray}
&&\mathcal{L}_2
= \mathcal{L}_1
+ \alpha [\partial_t S + \grad \cdot (S\mathbf{v}_n)]
+ \gamma [\partial_t (\beta S) + \grad \cdot (\beta S \mathbf{v}_n)].
\end{eqnarray}
with the two constraints corresponding to conservation of entropy and conservation of relative fluid vorticity. Now on the hydrodynamic length scales, the quantum pressure term appearing in Eq.\ (\ref{salman_eqn_Lagrangian1}) can be neglected. In this limit we can recover Landau's two fluid model as described in \cite{coste_98}.
    
\section*{Acknowledgments}

HS and NGB would like to acknowledge funding from the Isaac Newton Trust (Minute 8.10(a)) for supporting this work. \\

\bibliographystyle{ws-rv-van}
\bibliography{chapter.bib}

\end{document}